\documentstyle[12pt]{article}
\makeatletter                    
\@addtoreset{equation}{section}  
\makeatother                     

\begin{document}
\begin{titlepage}
\begin{flushright}
\today \ / \ HD-TVP-95-10
\end{flushright}
\vskip 3cm
\begin{center}
{\Large\bf Spin Densities in Pseudo-Classical Kinetic Theory
\footnote{Supported in part by
the Deutsche Forschungs Gemeinschaft, grant number Hu 233/4-3.}}
\vskip 1cm
\mbox{F.M.C.\ Witte} \\
{
\it
Institut f\"ur Theoretische Physik,\
Universit\"at Heidelberg\\
Philosophenweg  19, \ 69120  Heidelberg \\
Germany}
\end{center}
\vskip 1.5cm
\begin{abstract}
\footnotesize
It is shown that classical many-particle systems based on N=1 supersymmetry
allow for observable consequences of the spin degrees of freedom. In contrast
to the one-particle system, where a consistent formulation of spinspace density
is impossible, the many-particle system allows for spin to enter into the
equations of motion in a non-trivial way.
This density can then be directly compared to the decomposition of the
Wigner-operator in terms of spin matrices.
We discuss the quantization of a classical kinetic theory for N=1 particles,
both in the non-relativistic and the relativistic context.
{}From an expansion of the Dirac-spinors in terms of large and small components
it is seen that in the non-relativistic limit the pseudo-scalar, the time-like
component of the axial-vector current and the spatial components of the
vector-current vanish. The spatial components of the axial vector-current
vanish in the classical limit. The classical appearance of spin is due to the
spin-tensor contribution.

\end{abstract}

\end{titlepage}

\section{MOTIVATION}

\par
In recent years the prospect of identifying a quark-gluon plasma in
nucleus-nucleus or heavy-ion collisions has aroused great interest in the
theoretical description of chiral symmetry restoration under non-equilibrium
conditions.
Much of the effort in understanding the basic physics of this problem  has gone
into the formulation of a transport theory for the relevant degrees of freedom,
quarks, gluons and mesons. Traditionally these transport theories have been set
up for on-shell particles within the framework of a semi-classical
approximation.
By the methods of field theory similar equations can be derived \cite{chir2}
although the physical interpretation then remains partly obscured due to the
difficulty of interpreting the classical limit of quantum field theory in terms
of particles. The classical field is dominated by coherent states with an {\it
indefinite} particle number \cite{clas}.
One way out is to stick to field theory concepts and to try to go beyond the
semi-classical and on-shell approximations \cite{veld}. In this paper we rather
do the reverse.
\par
The details of the transition from classical to quantum physics, apart from its
formal mathematical context, are in general not well understood \cite{Hu}.
In particular, and relevant within the context of this work, there seems to be
no classical analogue of the quantum mechanical spin of fundamental fermions,
like quarks and leptons.
This should be contrasted with the orbital angular momentum, and even the spin
1 of vector bosons, which, apart from their quantized nature, can be accounted
for classically.
Although this is an intriguing problem by itself, it will not be of concern
here. Instead we will focus on a closely related question.
\par
In transport theories for relativistic fermions the classical limit appears to
contain a spin density \cite{spin2}.
As spin is considered to be a purely quantum mechanical effect this state of
affairs is at least confusing.
The goal of this paper is to remove this conceptual problem and analyze this
`classical' appearance of spin in detail.
We hope that a proper treatment of this problem eventually improves our
understanding of the $\hbar \rightarrow 0$ limit of relativistic quantum
transport theories for fermions.
\par
In this paper we treat two questions, closely related to eachother. First of
all we will discuss the role of spin in classical systems. In section 2, we
will introduce to this end the formalism of N=1 supersymmetric classical
mechanics. It will be demostrated to describe classical particles with spin and
it will be shown that this spin is unobservable in the 1-particle system.
In section 3, we will reconsider the question of measurabillity in a
many-particle system and set up a transport theory including the spin degrees
of freedom. At this point, supersymmetry is explicitly broken. A supersymmetric
kinetic equation is discussed elsewhere \cite{witte1}.
Second and last we will disentangle the non-relativistic and classical limiting
procedures in the quantum mechanical appearance of spin.
We discuss the spinor-decomposition of quantum Wigner functions in both
relativistic and non-relativistic settings.
We show there is no simple Foldy-Wouthuysen transformation that will yield the
non-relativistic Wigner-function when applied to the relativistic free theory.
An expansion of the relativistic Wigner-function in terms of large and small
components of the Dirac-spinors allows an analysis of the non-relativistic
limit. It can be seen that in the classical limit the axial-vector contribution
vanishes and the spin-tensor contributions survive. Pseudo-scalar contrinutions
vanish due to the non-relativistic limit.
Finally we summarize the main conclusions.

\section{N=1 Supersymmetric Classical Mechanics}
\par
The purpose of this section is to introduce the main tool used in this
analysis; N=1 supersymmetric mechanics.
We will start out by assigning both commuting and anti-commuting coordinates to
a single particle.
Next we write down a supersymmetric action principle that yields the free
particle equations of motion for the commuting coordinates.
After explicitly demonstrating that a particular bilinear form of the
anti-commuting coordinates represents an intrinsic angular momentum, we will
show that it is unobservable.
Since this section deals with a rather well documented system, our presentation
will resemble earlier works \cite{susy}.
\par
Let us introduce a super-time variable consisting of a pair $(t,\tau)$ of which
$t$ is a commuting and $\tau$ satisfies $\tau^2 = 0$.
A particle's position is specified by a commuting 3-vector ${\vec{X}}(t,\tau)$
which has the decomposition
\begin{equation}
\label{com}
 {\vec{X}}(t,\tau) = {\vec{x}}(t) + {\vec{\theta}}(t) \tau \ ,
\end{equation}
because a Taylor-expansion in $\tau$ truncates after the first order.
Since ${\vec{X}}$ is a commuting object, so is ${\vec{x}}$ whereas
${\vec{\theta}}$ must be anti-commuting.
A small translation in $(t,\tau)$ space has the following effect on ${\vec{X}}$
\begin{equation}
{\vec{X}}(t+\delta,\tau+\epsilon) = {\vec{X}}(t,\tau) +
\delta \frac{\partial}{\partial t} {\vec{X}}(t,\tau) +
\epsilon {\vec{\theta}}(t) \ .
\end{equation}
The generators of these `super-translations' are
\begin{equation}
{\cal{Q}} = \tau \frac{\partial}{\partial t} - \frac{\partial}{\partial \tau} \
\ and  \ \  H =
\frac{\partial}{\partial t} \ ,
\end{equation}
and they span the algebra
\begin{equation}
[{\cal{Q}},H] = [H,H] = 0 \ , \ [{\cal{Q}},{\cal{Q}}] = 2H \ .
\end{equation}
The brackets in this equation are super-commutators, i.e. they are commutators
(also denoted $[A,B]_{-}$) when
atleast one of the entries is commuting, and they are anti-commutators
($[A,B]_{+}$) when both entries are anti-commuting.
The transformation rules for the components of ${\vec{X}}$ are given by
\begin{equation}
\label{rules}
\delta {\vec{x}} = \alpha {\vec{\theta}} \ , \ \delta {\vec{\theta}} = \alpha
{\dot{\vec{x}}} \ ,
\end{equation}
where $\alpha$ is infinitesimal.
If ${\cal{P}}$ is the parity transpormation then we will assume that
${\vec{\theta}}$ transforms like a vector under parity. The reason for doing so
is obvious from Eq.(\ref{rules}). The supersymmetry transformation mixes
${\vec{\theta}}$ and ${\vec{x}}$ and so giving them different parity would lead
to vector and axial-vector component mixing. This is undesirable.
The quantity
\begin{equation}
\theta^{4} = \theta^{1}\theta^{2}\theta^{3} = \frac{1}{6}
\epsilon_{abc}\theta^{a}\theta^{b}\theta^{c}
\end{equation}
is easilly seen to be a pseudo-scalar.
\par
If we want to construct a supersymmetric action functional we need to use
derivatives $D$ that are covariant with respect to these translations,
that is $D$ must satisfy
\begin{equation}
[{\cal{Q}},D]_{+} = 0 \ .
\end{equation}
It is straightforward to check that
\begin{equation}
D = \tau \frac{\partial}{\partial t} + \frac{\partial}{\partial \tau}
\end{equation}
satisfies this requirement.

\subsection{Free Particles}

\par
The general form of a supersymmetric action functional is
\begin{equation}
S[{\vec{X}}] = \int dt d\tau \ {\cal{L}}({\vec{X}}) \ ,
\end{equation}
where the integration is standard over the
`normal' time $t$ {\it and} Berezin-integration
over the anti-commuting variable $\tau$.
Due to the fact that the integration measure $dtd\tau$
is anti-commuting, the minimal number of $D$'s in a non-trivial Lagrangian is
three.
The simplest non-trivial choice therefor is
\begin{equation}
S_{1}[{\vec{X}}(t,\tau)] = \int dt d\tau \ \frac{1}{2}
DX_{a}D(DX_{a}) \ ,
\end{equation}
where we introduced the latin indices $a=1,2,3$.
By working out the two factors
\begin{equation}
DX_{a}(t,\tau) = - \theta_{a}(t) + \tau {\dot{x}}_{a}(t) \ ,
\end{equation}
and
\begin{equation}
DDX_{a} = {\dot{\theta}}_{a}(t) \tau + {\dot{x}}_{a}(t) \ ,
\end{equation}
one finds after performing the Berezin-integration over $\tau$
\begin{equation}
S_{1} = \int dt \frac{1}{2} \{ {\dot{x}}_{a}{\dot{x}}_{a} + \theta_{a}
{\dot{\theta}}_{a} \} \ .
\end{equation}
The equations that follow from extremizing this action are
\begin{eqnarray}
{\ddot{x}}_{a} & = & 0 \nonumber \\
{\dot{\theta}}_{a} & = & 0 \nonumber \ , \\
\end{eqnarray}
indeed, for ${\vec{x}}$, the free particle equations of motion.

\subsection{Interaction with an External Field}

\par
In order to obtain insight in the physical content of the
anti-commuting variables ${\vec{\theta}}$ we add the following interaction
term to the action,
\begin{equation}
V = \frac{1}{2} \epsilon_{abc}B_{a}\theta_{b}\theta_{c} \tau \ , \ B=const.
\end{equation}
This interaction term breaks supersymmetry;
it leaves the equation for ${\vec{x}}$ unaffected while for the
${\vec{\theta}}$
we find
\begin{equation}
\label{eqmt}
{\dot{\theta}}_{a} = \epsilon_{abc}B_{b}\theta_{c} \ .
\end{equation}
The solution to Eq.(\ref{eqmt}) represents a vector precessing around the
${\vec{B}}$-axis with a frequency $\mid B \mid$. Ofcourse this looks like the
precession of an angular momentum in a homogeneous magnetic field.
To identify this angular momentum in detail we allow ${\vec{B}}$ to be
${\vec{x}}$-dependent.
In this case the equation for ${\ddot{x}}_{a}$ is modified to read
\begin{equation}
\label{eqm}
{\ddot{x}}_{a} = -
\nabla_{a}(\frac{1}{2}\epsilon_{bcd}B_{b}\theta_{c}\theta_{d}) =
-\nabla_{a}B_{b}S_{b} \ ,
\end{equation}
where we defined the vector ${\vec{S}}$ by
\begin{equation}
\label{spin}
S_{a} = \frac{1}{2} \epsilon_{abc}\theta_{b}\theta_{c} \ .
\end{equation}
Obviously ${\vec{S}}$ acts as an effective magnetic dipole in the
${\vec{B}}$-field. For ${\vec{S}}$ to be a proper angular momentum it must
satisfy the $SO(3)$ commutation relations.
Toc heck this we insert the expression for ${\vec{S}}$ into the supersymmetric
Poissonbrackets defined as
\begin{equation}
[f(\theta_{a}),g(\theta_{b})] = (\frac{\partial}{\partial
\theta_{a}}_{R}f(\theta))(\frac{\partial}{\partial \theta_{a}}_{L} g(\theta)) \
,
\end{equation}
where the subscripts $L$ and $R$ denote left- and right-derivative
respectively.
The vector defined in Eq.(\ref{spin}) indeed satisfies the SO(3) commutation
relations under poissonbracketing
\begin{equation}
[S_{a},S_{b}] = \epsilon_{abc}S_{c} \ ,
\end{equation}
identifying it as an angular momentum.
Using Eq.(\ref{spin}) we obtain for the time derivative of the spinvector
\begin{equation}
{\dot{S}}_{a} = \epsilon_{abc}{\dot{\theta}}_{b}\theta_{c} \ .
\end{equation}
When we substitute Eq.(\ref{eqmt}) in the previous equation, and use the fact
that $\theta^2 = 0$, we find the equation of motion for ${\vec{S}}$ to read
\begin{equation}
{\dot{S}}_{a} = \epsilon{_abc}B_{b}S_{c} \ .
\end{equation}
We can complete the algebra by denoting that
\begin{equation}
[ \theta_{a},\theta_{b} ] = \delta_{ab} \ ,
\end{equation}
and
\begin{equation}
[ \theta_{a},S_{b}] = \epsilon_{abc}\theta_{c} \ .
\end{equation}
So we conclude that the N=1 supersymmetric particle is infact a classical
particle with an intrinsic angular momentum, i.e. spin. In particular, one can
see from its definition Eq.(\ref{spin}) that spin is a axial vector, i.e. under
the parity operation ${\cal{P}}$ we have for the vectors
${\vec{x}},{\vec{\theta}}$ and ${\vec{S}}$ that
\begin{equation}
{\cal{P}} \{ {\vec{x}},{\vec{\theta}},{\vec{S}} \} = \{ - {\vec{x}}, -
{\vec{\theta}},{\vec{S}} \} \ .
\end{equation}
Note that this fixes ${\vec{B}}$ as an axial vector aswell.

\subsection{Measurement on anti-commuting Quantities}

\par
Since all measurements yield real numbers, the existence of an experiment that
measures some effect of the anti-commuting degrees of freedom is closely linked
to the existence of a map ${\cal{F}}$ that maps ${\vec{\theta}}$ into the real
numbers. In practice this boils down to some kind of averaging over the
anti-commuting degrees of freedom
\begin{equation}
{\cal{F}}:\ < g > = \int d^{3}\theta \ g({\vec{\theta}}) f({\vec{\theta}},t) \
, \end{equation}
with some weight-function $f$.
We will define the measure $\int d^3 \theta$ is a scalar under parity
transformations. By this choice we deviate from standard notations, but we
belive our convention is more natural. The reason for doing so is twofold, on
the one hand, upon quantization Berezin-integration goes over into the $Tr$
operation of taking traces. The latter is obviuosly a scalar under parity.
On the other hand we do not want the transformation properties of $\int d^3
\theta$ to interfere with those of $\int d^3 x d^3 p$.
So $\int \ d^3x d^3 \theta \ \theta_{4}$ is a pseudo-scalar.
For the sake of consistency, only commuting objects should generate a
non-vanishing average. This constraint on $f$ implies it is of the form
\begin{equation}
\label{f1}
f(\theta_{a}) = \theta_{a}\theta_{b}\theta_{c} + \frac{1}{2} C_{a}\theta_{a} \
{}.
\end{equation}
The first term here allows pure c-numbers to be equal to their average.
The second term yields an average value for the spin vector ${\vec{S}}$ by
\begin{eqnarray}
< S_{a} > & = & \int d^{3}\theta \ \epsilon_{abc}\theta_{b}\theta_{c}
C_{d}\theta_{d} \nonumber \\
 & = & \frac{1}{2} \epsilon_{abc}\epsilon_{bcd}C_{d} < \theta_{4} > = C_{a} <
\theta_{4} > \nonumber \ .\\
\end{eqnarray}
Obviously ${\vec{C}}$ must be a vector, since ${\vec{S}}$ is an axial vector.
Since we have chosen ${\vec{\theta}}$ to be a vector in the introduction to
this section, we see that ${\cal{F}}$ is a scalar.
\par
An additional requirement is \cite{susy}
\begin{equation}
\label{req}
<g({\vec{\theta}})g^{*}({\vec{\theta}})> \geq 0 \ .
\end{equation}
Yet by inserting the functions
\begin{equation}
g_{\pm}(\theta_{a}) = \theta_{1} \pm \imath \theta_{2} \ ,
\end{equation}
we obtain
\begin{equation}
<g_{\pm}g_{\pm}^{*}> = \mp 2 C_{3} \ .
\end{equation}
Depending on the sign of $C_{3}$ Eq.(\ref{req}) fails either for $g_{+}$ or for
$g_{-}$.
Choosing ${\vec{C}}=0$ entirely trivializes $f(\theta_{a})$,
so we must conclude that in the 1-particle system no experiment, i.e.
non-trivial $f({\vec{\theta}})$, can detect the presence of the anti-commuting
degrees of freedom.
In particular this means, by using Eq.(\ref{f1}) with ${\vec{C}}=0$ in
Eq.(\ref{eqm}), that
\begin{equation}
\label{sad}
< {\ddot{x}}_{a}> = 0 \ .
\end{equation}
The interaction of ${\vec{B}}$ with ${\vec{S}}$ causes no observable effect on
the equations of motion of the particle.
\par
This may at first sight seem an unavoidable consequence of
including degrees of freedom which have an `unphysical'
anti-commuting nature. Yet upon quantization it can be shown that
Eq.(\ref{req}) {\it can} be fulfilled \cite{susy}.
We will postpone any discussion of this effect to the final section where we
treat a quantized system.
In the next section we set out to show that the consequences of Eq.(\ref{sad})
can be avoided. Many-particle systems will allow for observable consequences of
spin.

\section{N=1 Many-Particle System}
\par
In this section, we will investigate the properties of the spin-vector
${\vec{S}}$ defined in the previous section, in a many-particle system.
In the case of orbital angular momentum the limit $\hbar \rightarrow 0$ implies
that only large quantum numbers will survive. In the case of spin this is
obviously no remedy. Yet, from physical experience we know that systems
containing an extremely large {\it number} of spins allow for observable
consequences of the interactions among the spins.
So it seems natural to consider a many-particle system as a possible way to
study a `classical' appearance of spin.
Furthermore, as we will now show, in a two-particle system one can satisfy
Eq.(\ref{req}).

\subsection{2 particle system}

\par
The main problem with the density $f$ in the 1-particle system was its
inabillity to handle complex functions of the $\theta_{a}$ correctly.
A two particle system offer the opportunity to evade this problem at the
expense of restricting the possible values for the total spin.
Consider a distribution function of the form
\begin{equation}
\label{twof}
f({\vec{\theta}}^{1},{\vec{\theta}}^{2}) =
\theta_{a}^{1}\theta_{b}^{1}\theta_{c}^{1}
\theta_{d}^{2}\theta_{e}^{2}\theta_{f}^{2} +
C_{a}\theta_{a}^{1}\theta_{b}^{2}\theta_{c}^{2}\theta_{d}^{2} +
D_{a}\theta_{a}^{2}\theta_{b}^{1}\theta_{c}^{1}\theta_{d}^{1}  \ .
\end{equation}
When taking the average
\begin{equation}
<g({\vec{\theta}}^{1})g^{*}({\vec{\theta}}^{1}) +
g({\vec{\theta}}^{2})g^{*}({\vec{\theta}}^{2}) > = \mp \{ C_{3} + D_{3} \} \ ,
\end{equation}
we see that Eq.(\ref{req}) can be satisfied if the total spin vanishes and if
we restrict our attention to averages.
We are thus lead to the formulation of measurabillity in a statistical sense.
In particular by using the delta functions
\begin{equation}
\int \delta ( \theta - \theta^{i} ) d\theta = 1 \ , \ \int \theta \delta (
\theta - \theta^{i} ) d\theta = \theta^{i} \ , \ \delta^{3} ( {\vec{\theta}} -
{\vec{\theta}}^{i} ) = \delta ( \theta_{a} - \theta^{i}_{a})\delta ( \theta_{b}
- \theta^{i}_{b})\delta ( \theta_{c} - \theta^{i}_{c})
\end{equation}
we can rewrite Eq.(\ref{twof}) as a two particle spinspace density $F_{2}$
\begin{equation}
F_{2}({\vec{\theta}}) =
\label{ff}
\prod_{i=2}^{2} \delta^3({\vec{\theta}} - {\vec{\theta}}^{i}) + \sum_{i=1}^{2}
C_{a}^{i}\theta_{a}\delta^3({\vec{\theta}}-{\vec{\theta}}^{i}) \prod_{j \neq i}
\theta_{b}^{j}\theta_{c}^{j}\theta_{d}^{j} \ ,
\end{equation}
and averages are calculated from
\begin{equation}
<g({\vec{\theta}})> = \int d^3 \theta \{\int \prod_{i}d^3 \theta^{i}
g({\vec{\theta}}) F_{2}({\vec{\theta}}) \} \ .
\end{equation}
These averages now behave fine.
\par
A direct consequence of this reformulation is that the equations of motion
for ${\vec{x}}$ become non-trivial
\begin{equation}
\label{2eqm}
{\ddot{x}}_{a}^{i} = - \nabla_{a} (\frac{1}{2} \epsilon_{bcd} B_{b}
\theta_{c}^{i} \theta_{d}^{i}) = -\nabla_{a}B_{b}S_{b}^{i} \ , \ i=1,2\ .
\end{equation}
\par
We have achieved that the averaging procedure is now well-behaved with respect
to linear complex functions of ${\vec{\theta}}$. Furthermore we notice that the
anti-commuting variables enter the equations of motion for ${\vec{x}}$ only
through their quadratic combination in ${\vec{S}}$.
In the next subsection we will see how this can be exploited.
The basic idea is to generalize to an $M$-particle system, $F_{M}$,
include the commuting dergrees of freedom and show that the density obtained in
this way satisfies a Klimontovich equation.
It can then be identified with an {\it exact} phase/spin space density by
reexpressing all the dependence on ${\vec{\theta}}$ in terms of the spin
${\vec{S}}$.
Suitably averaging this exact $M$-particle density then yields a kinetic
equation
for the averaged phase/spin space density.
In the limit $M \rightarrow \infty$ we then obtain a Vlasov equation.

\subsection{The N=1 Many-particle System}
\par
The purpose of this subsection is to show that for a supersymmetric
$M$-particle system an exact phase/spin space density satisfying a
Liouville-type  evolution equation can be constructed.
The $M$-particle generalization of Eq.(\ref{ff}), including the commuting
degrees of freedom, is given by
\begin{equation}
\label{MF}
F_{M}({\vec{x}},{\vec{p}},{\vec{\theta}};t) = \{ \sum_{i=1}^{M} \ \delta^{3}
({\vec{x}}-{\vec{x}}^{i}(t)) \delta^{3} ({\vec{p}}-{\vec{p}}^{i}(t)) \}
\{\prod_{i=1}^{M} \delta^3({\vec{\theta}} - {\vec{\theta}}^{i}) +
\sum_{i=1}^{M}
C_{a}^{i}(t)\theta_{a}\delta^3({\vec{\theta}}-{\vec{\theta}}^{i}) \prod_{j \neq
i}\theta_{a_{j}}^{j}\theta_{b_{j}}^{j}\theta_{c_{j}}^{j} \} \ ,
\end{equation}
as a function of the coordinate ${\vec{x}}$, momentum ${\vec{p}}$,
${\vec{\theta}}$ and time $t$.
The time dependence of $F_{M}$ originates from the particle coordinates which
depend on time, and from the time-dependent vectors ${\vec{C}}^{j}$.
In order to fix this time dependence we ressort to the so-called Klimontovich
equation for exact phase space densities.
Let $K$ be the super-Liouville operator defined by
\begin{equation}
K = \frac{\partial}{\partial t} +
{\dot{x}}_{a}\frac{\partial}{\partial x_{a}} +
{\dot{p}}_{a}\frac{\partial}{\partial p_{a}} +
{\dot{\theta}}_{a}\frac{\partial}{\partial \theta_{a}} \ ,
\end{equation}
then it is easy to show that $F_{M}$ satisfies the Klimontovich equation
\begin{equation}
\label{klim}
K F_{M}({\vec{x}},{\vec{p}},{\vec{\theta}};t) = 0 \ .
\end{equation}
Inserting Eq.(\ref{MF}) into the Klimontovich equation, Eq.(\ref{klim}),
yields the following expression
\begin{equation}
\label{eqC}
\sum_{i=1}^{M}
{\dot{C}}_{a}^{i}(t)\theta_{a}\delta^3({\vec{\theta}}-{\vec{\theta}}^{i})
\prod_{j \neq i} \theta_{a_{j}}^{j}\theta_{b_{j}}^{j}\theta_{c_{j}}^{j} =
\sum_{i=1}^{M}C_{a}^{i}(t){\dot{\theta}}_{a}
\delta^3({\vec{\theta}}-{\vec{\theta}}^{i}) \prod_{j \neq i}
\theta_{a_{j}}^{j}\theta_{b_{j}}^{j} \theta_{c_{j}}^{j} \ ,
\end{equation}
from which to solve for ${\vec{C}}^{j}(t)$.
Equating each summand seperately and using Eq.(\ref{eqmt}) gives
\begin{equation}
{\dot{C}}_{a} = \epsilon_{abc}B_{b}C_{c} \ ,
\end{equation}
the equation of motion for the vectors ${\vec{C}}^{j}$.
It coincides with the equation for the spinvector ${\vec{S}}$,
as expected.
When averaging Eq.(\ref{eqC}) over all ${\vec{\theta}}^{i}$ one obtains
\begin{equation}
\sum_{i=1}^{M}{\dot{C}}_{a}(t)^{i}\theta_{a} =
\sum_{i=1}^{M}C_{a}(t)^{i}{\dot{\theta}}_{a} \ ,
\end{equation}
implying that when the total spin vanishes,
it remains zero during the entire time-evolution.
\par
We have seen that $F_{M}({\vec{x}},{\vec{p}},{\vec{\theta}};t)$
is an exact phase/spin space density satisfying an evolution equation
Eq.(\ref{klim}). Yet, the dynamics of the system is still rather simple and
will not lead to an interesting kinetic equation because the interactions among
the particles is missing.
Improving this state of affairs will require us to introduce an
${\vec{x}}$-dependent vectorfield ${\vec{B}}$.
Furthermore the vectors ${\vec{C}}^{j}$ will then also become dependent on the
particle-position and hence the spin-density, i.e. the second term in
Eq.(\ref{MF}), will become a local quantity.
Consequently we would once more face the problem of satisfying Eq.(\ref{req}),
knowing that it will be violated locally anyhow.
However, now the situation is different. We are working in a many-particle
enviroment and we will proceed towards a statistical description of the system.
In particular, the question of the measurabillity of the coordinates
${\vec{\theta}}$ is no longer of interest since it represents microscopic
information. The relevant physical observable relating to the anti-commuting
degrees of freedom has now become the macroscopic expectation value of
${\vec{S}}$.

\subsection{ N=1 Transport Theory}
\par
In this section we want to take the final step in our argument. We will
assume some kind of averaging of the exact density $F_{M}$ and show how the
resulting smoothed density $f$ satisfies a transport equation.
Let us decompose $F_{M}$ into an averaged part and a fluctuation part as
\begin{equation}
\label{av}
F_{M}({\vec{x}},{\vec{p}},{\vec{\theta}};t) =
<F_{M}({\vec{x}},{\vec{p}},{\vec{\theta}};t)>_{av} + \delta
F_{M}({\vec{x}},{\vec{p}},{\vec{\theta}};t) \ .
\end{equation}
The exact nature of the averaging is immaterial, one should note however that
any averaging over the anti-commuting ${\vec{\theta}}^{i}$ will remove all the
dependence on these variables due to the nature of Berezin-integration.
So the function
\begin{equation}
f({\vec{x}},{\vec{p}},{\vec{\theta}};t) =
<F_{M}({\vec{x}},{\vec{p}},{\vec{\theta}};t)>_{av}
\end{equation}
will only depend on those spins, through their ${\vec{C}}^{j}$, lying within
the volume-elements of phase-space averaged over.
\par
We add particle interactions by relating the vectorfield ${\vec{B}}$ with
$f$ through the electrodynamics relation
\begin{equation}
B_{a}({\vec{x}},t)^{av} = \frac{\mu_{0}}{2 \pi} \int d^{3}x'
\frac{\epsilon_{abc}\epsilon_{bde} \{\nabla_{d} M_{e}({\vec{x}}')^{av}\}(x_{c}
- x_{c}')}{ \mid{\vec{x}} - {\vec{x}}'\mid^{3}} \ ,
\end{equation}
where the magnetisation ${\vec{M}}$ is given by
\begin{equation}
M_{a}({\vec{x}})^{av} = \int d^3{\vec{p}} d^3{\vec{\theta}}
S_{a}f({\vec{x}},{\vec{p}},{\vec{\theta}};t) \ .
\end{equation}
{}From Eq.(\ref{av}) we see that the exact quantities are related to their
averages by adding fluctuations
\begin{eqnarray}
{\vec{B}} & = & {\vec{B}}^{av} + \delta {\vec{B}} \nonumber \\
{\vec{M}} & = & {\vec{M}}^{av} + \delta {\vec{M}} \nonumber \\ \ .
\end{eqnarray}
By observing that
\begin{equation}
\frac{\partial}{\partial \theta_{b}} S_{a} = 2 \epsilon_{abc} \theta_{c} \ ,
\end{equation}
implies
\begin{equation}
{\dot{\theta}}_{b}\frac{\partial}{\partial \theta_{b}} S_{a} = {\dot{S}}_{a} \
,
\end{equation}
we may interpret $f({\vec{x}},{\vec{p}},{\vec{\theta}};t)$ as a function of
${\vec{S}}$ and use
\begin{equation}
{\dot{\theta}}_{b}\frac{\partial}{\partial \theta_{b}}
f({\vec{x}},{\vec{p}},{\vec{\theta}};t) = {\dot{S}}_{b}\frac{\partial}{\partial
S_{b}} f({\vec{x}},{\vec{p}},{\vec{S}};t) \ .
\end{equation}
Since the quantity ${\vec{S}}$ is commuting and appears at most linearly in all
expressions we replave it by its c-number representation, its expectation
value.
The spin-dependent term in $F_{M}$ is now
\begin{equation}
\sum_{i=1}^{M} \ \delta^3 ({\vec{x}}-{\vec{x}}^{i}(t)) \delta^3
({\vec{p}}-{\vec{p}}^{i}(t))\delta^3 ({\vec{S}}-{\vec{C}}^{i}(t)) \ .
\end{equation}
However, the spin-independent term must be annihilated when calculating the
total spin and thus is proportional to
\begin{equation}
\sum_{i=1}^{M} \ \delta^3 ({\vec{x}}-{\vec{x}}^{i}(t)) \delta^3
({\vec{p}}-{\vec{p}}^{i}(t))\delta^3 ({\vec{S}}) \ .
\end{equation}
All reference to the anti-commuting variables has dissappeared.
Averaging the Klimontovich equation Eq.(\ref{klim}) now yields
\begin{eqnarray}
\label{avklim}
\{ \frac{\partial}{\partial t} +
{\dot{x}}_{a}\frac{\partial}{\partial x_{a}} -
\frac{\partial(B_{a}^{av}M_{a}^{av})}{\partial x_{a}}\frac{\partial}{\partial
p_{a}} & + & (\epsilon_{abc}B_{b}^{av} M_{c}^{av}) \frac{\partial}{\partial
S_{a}} \}f({\vec{x}},{\vec{p}},{\vec{S}};t) \nonumber \\
& = &  \frac{\partial(\delta B_{a} \delta M_{a})}{\partial
x_{a}}\frac{\partial}{\partial p_{a}} + (\epsilon_{abc}\delta B_{b} \delta
M_{c})\frac{\partial}{\partial S_{a}} \}\delta
F({\vec{x}},{\vec{p}},{\vec{S}};t) \nonumber \ .\\
\end{eqnarray}
The collisions term can be extracted from the r.h.s. of this equation.
If we assume that as $M \rightarrow \infty$ the fluctuations can be neglected
and if we remove the explicit notation, $< \ >_{av}$, from the averaged
quantities we find
\begin{equation}
\label{eqki}
\{ \frac{\partial}{\partial t} +
{\dot{x}}_{a}\frac{\partial}{\partial x_{a}} -
\frac{\partial({\vec{B}}.{\vec{M}})}{\partial x_{a}}\frac{\partial}{\partial
p_{a}} + ({\vec{B}} \wedge {\vec{M}}) \frac{\partial}{\partial S_{a}}
\}f({\vec{x}},{\vec{p}},{\vec{S}};t) = 0 \ ,
\end{equation}
the Vlasov equation for the system.
Eq.(\ref{eqki}) describes the transport phenomena that take place in this
many-particle system of particles with spin due to mutual spin-spin
interactions. The collission term can be retrieved from the Klimontovich
equation by giving the righ hand side of Eq.(\ref{avklim}) a more detailed
treatment \cite{chir2}.

\section{Quantum spin}
\par
In the previous sections we have established a {\it classical} kinetic theory
explicitly containing spin degrees of freedom. The result of our labor was an
equation, Eq.(\ref{eqki}), describing the non-equilibrium physics of a system
containing a very large number of particles, each carrying a magnetic dipole
moment proportional to its spin. In this section we make contact with the
relativistic formulation of quantum kinetic theory for particles with spin.
We proceed, by formulating a decomposition of the phase/spin space density
that upon quantization grows into the spinor-decomposition of the fermionic
Wigner-function \cite{spin2}. The latter is then expanded in terms of large and
small components allowing a carefull seperation between the non-relativistic
and the classical limit.

\subsection{Non-Relativistic Spin}
\par
If we write down the most general, internally consistent, expansion of the
phase-spin space density in terms of products of ${\vec{\theta}}$, one finds
\begin{equation}
\label{pseudowig}
f = s\theta_{a}\theta_{b}\theta_{c} + p\theta_{4} + {\vec{C}}.{\vec{\theta}} \
{}.
\end{equation}
Let us discuss the terms not appearing in this expansion.
We could have added a term proportional to
\begin{equation}
T_{ab}\theta_{a}\theta_{b} + \frac{1}{2}t_{a}
\epsilon_{abc}\theta_{a}\theta{b}\ ,
\end{equation}
for some vector ${\vec{t}}$ and some anti-symmetric tensor $T_{ab}$. They will
generate an expectation value for ${\vec{\theta}}$. The vector ${\vec{t}}$ will
give an axial vector contribution to $<{\vec{\theta}}>$, but since this is a
vector we must have ${\vec{t}}=0$. The tensor $T_{ab}$ will give a vector-like
contribution and therefor seems to be acceptable. But remember that
${\vec{\theta}}$ is an anti-commuting quantity.
So any consistently defined average value would have to satisfy
\begin{equation}
<\theta_{1}\theta_{2}> = - <\theta_{2}\theta_{1}> \ .
\end{equation}
But if we rewrite this in terms of connected and disconnected contributions we
find
\begin{equation}
<\theta_{1}\theta_{2}>_{con} + <\theta_{2}\theta_{1}>_{con} =
2<\theta_{1}><\theta_{2}> \ .
\end{equation}
This equation can only be true for $<{\vec{\theta}}>=0$, and thus
$T_{ab}=0$.
Finally an additional linear contribution of the form
\begin{equation}
\epsilon_{abc}a_{ab}\theta_{c}
\end{equation}
for some anti-symmetric tensor $a_{ij}$ will give a vector-like contribution
to the average of ${\vec{S}}$, which is axial. So $a_{ij}=0$.
We see that Eq.(\ref{pseudowig}) is infact the most general expansion we can
make.
\par
Now consider the quantization of the 1-particle system.
Classically the anti-commuting coordinates ${\vec{\theta}}$ satisfy the
following Poissonbracket
\begin{equation}
[\theta_{a},\theta_{b}]_{P} = \delta_{ab} \ .
\end{equation}
Quantization now implies that we make the transition to the {\it
anti}-commutator
\begin{equation}
[\theta_{a},\theta_{b}]_{+} = \imath \hbar  \delta_{ab} \ .
\end{equation}
By defining
\begin{equation}
\sigma_{a} = \imath \sqrt{\frac{2}{\imath \hbar}}\theta_{a} \ ,
\end{equation}
the components of ${\vec{\tau}}$ now satisfy the anti-commutation relations
\begin{equation}
[\sigma_{a},\sigma_{b}]_{+} =  2 \delta_{ab} \ ,
\end{equation}
defining a Clifford algebra.
Note that ${\vec{\sigma}}$ transforms like a vector under parity, hence they
are not generators of $SO(3)$ or one of its representations and cannot be
identified with the Pauli spinmatrices. We will come back to this point
shortly.
The ${\vec{\sigma}}$ can be identified with a set of two by two matrices and
substituting them into the 1-particle spindensity, Eq.(\ref{f1}), yields
\begin{equation}
\label{qspin}
f({\vec{\theta}}) =  ( 1 + {\vec{C}}.\frac{2 \imath {\vec{\sigma}}}{\hbar} ) \
,
\end{equation}
within a factor $(\frac{\hbar}{2 \imath})^{3/2}$. The integration over the
anti-commuting variables that yielded averages in the pseudo-classical limit is
now replaced by taking traces over the spin-indices. In particular, for the
spin-operator quantization yields
\begin{equation}
S_{a} = \epsilon_{abc}\sigma_{b}\sigma_{c} =
\frac{1}{2}\epsilon_{abc}[\sigma_{b},\sigma_{c}]_{-} ,
\end{equation}
directly relating it to the commutator of $\sigma$-matrices.
The matrices $S_{a}$ we identify with the Pauli matrices ${\vec{\tau}}$ via
\begin{equation}
S_{a} = \tau_{a} \ .
\end{equation}
The anti-commutation relations of the ${\vec{\sigma}}$-matrices can be used to
show that ${\vec{S}}$, and thus ${\vec{\tau}}$, still satisfy the
$SO(3)$-commutation relations. By using that $(\sigma_{a})^2 = 1$ for any $a$
and projecting ${\vec{S}}$ on ${\vec{\sigma}}$ we find
\begin{equation}
\sigma_{a}\{\sigma_{b} S_{b} \} = \sigma_{a}
\epsilon_{bcd}\sigma_{b}\sigma_{c}\sigma_{d} = \sigma_{a} \sigma_{4} = \tau_{a}
\ ,
\end{equation}
which clearly displays the correspondence between the pseudo-classical averages
and quantum averages. The introduction of two sets of matrices may seem clumsy,
but consistency demands it. By using Eq.(\ref{qspin}) once again we get
\begin{equation}
<{\vec{S}}> = Tr \{{\vec{S}} f({\vec{\sigma}}) \} = {\vec{C}}<\sigma_{4}> \ .
\end{equation}
The quantum spin-density operator {\it can} satisfy our requirement,
Eq.(\ref{req}), provided
\begin{equation}
\mid {\vec{C}} \mid \leq \frac{1}{2} \hbar \ ,
\end{equation}
the equality, which is satisfied for both quarks and leptons, represents a pure
state.
\par
We now turn to the Wigner function $W({\vec{x}},{\vec{p}})$ defined by
\cite{kad}
\begin{equation}
\label{wig}
W_{\alpha \beta}({\vec{x}},{\vec{p}};t) = \int d^4y \  <
\psi_{\alpha}^{*}({\vec{x}} - \frac{{\vec{y}}}{2},t)
\psi_{\beta}({\vec{x}} + \frac{{\vec{y}}}{2},t) > \exp\{
\frac{\imath}{\hbar}{\vec{p}}.{\vec{y}} \} \ ,
\end{equation}
in terms of the non-relativistic field-operator $\psi_{\alpha}$, explicitly
including the spin-indices.
It is a 2 by 2 matrix in spin-space and can thus be decomposed in terms of the
generators of the algebra of 2 by 2 matrices.
These generators are
\begin{equation}
T_{i} = \{ \delta , \sigma^{4} = (\sigma^{1}\sigma^{2}\sigma^{3}) ,
{\vec{\sigma}} ,  \sigma^{4}{\vec{\sigma}} \} \ .
\end{equation}
We recognize the scalar, pseudo-scalar, vector and pseudo-vector contributions
in, respectively, $T_{1}$ , $T_{2}$, $T_{3-5}$ and $T_{6-8}$.
In this basis we can decompose $W$ as
\begin{equation}
\label{decom}
W_{\alpha \beta} = s \delta + p \sigma^{4} + {\vec{V}}.{\vec{\sigma}} +
{\vec{A}}\sigma^{4} {\vec{\sigma}} \ .
\end{equation}
Although the Wigner-function has only four independent components using an
eight dimensional basis for the expansion does not double this number. The
doubling comes from the extra splitting caused by including parity
transformations. Any function can be written as a sum of parity even and uneven
functions.
Through the relation between ${\vec{\sigma}}$ and the anti-commuting
coordinates ${\vec{\theta}}$ given by Eq.(\ref{qspin}) we can immediatly deduce
the naive pseudo-classical form of the distribution function
$f({\vec{x}},{\vec{p}},{\vec{\theta}};t)$
\begin{equation}
f = s` \theta_{a}\theta_{b}\theta_{c} + p'\theta_{4}  +
{\vec{V}}'.{\vec{\theta}} + {\vec{A}}'.\theta_{4}{\vec{\theta}} \ ,
\end{equation}
where the primes denote that these coefficients are only up to a factor equal
to those in Eq.(\ref{decom}).
Ofcourse we see that this exactly matches the decomposition found earlier in
Eq.(\ref{pseudowig}).
The axial vector, ${\vec{A}}'$, cannot yield a classical observable due to the
fact that for every component of the anti-commuting ${\vec{\theta}}$ we have
$\theta_{a}^2=0$.
So the axial vector in the Wignerfunction decomposition is a purely quantum
mechanical object and should vanish in the classical limit.
The generalization of these results to relativistic Wigner-operators for
fermions will be the goal of the final subsection. In particular we will see
that a pseudo-scalar contribution vanishes in the non-relativistic limit, so
that we can set $p=p'=0$ in the previous equations.

\subsection{Relativistic Spin}
\par
In the case of relativistic fermions the above treatment must be modified.
In this subsection we will discuss these modifications without going through
the whole derivations of the previous sections again.
In particular, we will focus on the decomposition of the Wigner operator, and
we will not discuss relativistic pseudo-classical kinetic equations.
\par
First of all we have to introduce an anti-commuting four vector $\theta^{\nu}$.
Together with the standard commuting coordinates $x^{\nu}$ it forms the
commuting object
\begin{equation}
X^{\nu}(\sigma , \tau) = x^{\nu}(\sigma) + \theta^{\nu}(\sigma) \tau \ ,
\end{equation}
where the pair $(\sigma , \tau)$ is now a super-worldline parameter.
Upon quantization the algebra of the anti-commuting coordinates becomes a
Clifford algebra and hence we obtain the identification
\begin{equation}
\theta^{\mu} \rightarrow \gamma^{\mu} \ .
\end{equation}
The pseudo-classical $\theta^{\mu}$ can be constructed from the
three-dimensional
coordinates ${\vec{\theta}}$ in exactly the same way as, in the quantum
mechanical case, the $\gamma^{\mu}$ are constructed from the ${vec{\sigma}}$.
If we seek a 4-dimensional generalization of ${\vec{S}}$ we find
\begin{equation}
S_{a} = \epsilon_{abcd}\theta_{b}\theta_{c} \ \rightarrow \
S_{\alpha \beta} = \epsilon_{\alpha \beta \gamma
\rho}\theta_{\gamma}\theta_{\rho} \ ,
\end{equation}
that it generalizes into an anti-symmetric 4-tensor, rather than into an axial
4-vector. Written out this gives
\begin{equation}
S_{\alpha \beta} = \left( \begin{array}{cc} 0 & {\vec{S}} \\ {\vec{S}} &
\epsilon_{abc} \theta_{0}\theta_{c} \\ \end{array} \right) \ .
\end{equation}
This is in contrast to the existing literature where spin is rather identified
with an axial 4-vector. A closer look at the relativistic wigner-function and
its non-relativistic limit will reveal the origin of this contradiction.

\subsubsection{Foldy Wouthuysen transformations and a small component
expansion}
\par
The non-relativistic limit of the Dirac equation can be found systematically
within the framework of Foldy-Wouthuysen transformations \cite{ytzi}. Let
$\Psi$ be a Dirac spinor given in terms of two two-spinor components $\Psi = (
\phi , \chi )$, then we define a unitary transformation
\begin{equation}
\Psi \rightarrow e^{-\imath Z} \Psi \ .
\end{equation}
The matrix $Z$ is now determined by the requirement that the new Hamiltonian
$H'$
\begin{equation}
H` =   e^{\imath Z} H e^{-\imath Z} \ ,
\end{equation}
no longer mixes the different two-spinor components. Physically this implies
that particle and anti-particle excitations decouple. Obviously only for the
free theory can we find an exact transformation of this type. In this case $Z$
is of the form
\begin{equation}
Z = -\imath {\vec{\gamma}}.{\vec{b}} \omega \ ,
\end{equation}
where ${\vec{b}}$ is a unit-vector. For all some interacting cases an
approximate Foldy-Wouthuysen transformation can be found for low-energy
fermions. A standard result from these considerations is that the relative
weight of particle and anti-particle excitations is given by
\begin{equation}
\phi \propto \frac{p}{m} \chi \ ,
\end{equation}
where $p$ is a typical momentum.
\par
Now consider the general form of the spinor-decomposition of the relativistic
Wigner function for Dirac-fermions
\begin{equation}
\label{relwig}
W = <{\bar{\Psi}} \Psi > = {\cal{F}} \delta + \imath \gamma^{5} {\cal{P}} +
{\cal{V}}_{\mu} \gamma^{\mu} + {\cal{A}}_{\mu} \gamma^{5}\gamma^{\mu} +
{\cal{S}}_{\mu \nu} \sigma^{\mu \nu} \ ,
\end{equation}
where we have scalar-, pseudo-scalar-, vector-, axial-vector, and
tensor-contributions to the Wigner-function. \{ We surpress all coordinate
dependencies from our notation since they are irrelevant for our purpose.\}
If we apply the previous Foldy-Wouthuysen transformation to the Wigner-function
we find that it is not sufficient to reduce Eq.(\ref{relwig}) to its
non-relativistic form. For example the time-like component of the axial-vector
current ${\cal{A}}^{0}$ is easilly seen to be invariant;
\begin{equation}
e^{\imath Z} \gamma^{5}\gamma^{0} e^{-\imath Z} \ = \gamma^{5} \gamma^{0} \ .
\end{equation}
However it is a $(\phi , \chi )$-mixing quantity and should therefore be
eliminated from the expressions.
The simplest and for our purposes sufficient method of finding the
non-relativistic limit is by explicitly introducing the large and small
components. We rewrite the Wigner-function in terms of $\phi$ and $\chi$
\begin{equation}
W = \left( \begin{array}{cc}  < \phi^* \phi > & < \phi* \chi > \\
- < \chi* \phi > & - < \chi* \chi > \\ \end{array} \right) = \left(
\begin{array}{cc}  W_{\phi} & W_{mix} \\
- W_{mix}^* & - W_{\chi} \\ \end{array} \right) \  ,
\end{equation}
or in terms of the block-matrices $W_{i}$. These can be explicitly computed in
terms of the coefficients appearing in the expansion Eq.(\ref{relwig}) by using
the Dirac-representation of the $\gamma^{\mu}$-matrices in terms of the
generators of the Clifford algebra ${\vec{\sigma}}$.
For the large components we get
\begin{equation}
W_{\phi} = \{ {\cal{F}} + {\cal{V}}^{0} \} \delta - {\cal{A}}_{i}
\sigma^{4}\sigma^{i} +
{\cal{S}}_{ij} \epsilon^{ijk} \sigma_{k} \ .
\end{equation}
Note that the scalar density appearing here is the sum of particle density \{=
particles + anti-particles\} and fermion-number density \{= particles -
anti-particles\}, in which the anti-particle contributions cancel out.
The pseudo-scalar is a $(\phi , \chi)$ mixing quantity and thus is surpressed
in the non-relativistic limit. If we now take the spin-operator defined in
terms of the matrices ${\vec{\sigma}}$ and calculate its average with
$W_{\phi}$ we find
\begin{equation}
\label{Srel}
< S_{a} > =  [ - {\cal{A}}_{i}Tr\{\sigma^{4}
\epsilon_{abc}\sigma_{b}\sigma_{c}\sigma_{i} \} + {\cal{S}}_{ij} \epsilon^{ijk}
Tr\{ \sigma_{k} \epsilon_{abc} \sigma_{b} \sigma_{c}\} ]  \ .
\end{equation}
Note that the first term reduces to
\begin{equation}
{\cal{A}}_{a}Tr\{(\sigma^{4})^2\} \ ,
\end{equation}
and the second to
\begin{equation}
\epsilon_{aij}{\cal{S}}_{ij} Tr\{\epsilon_{bcd}\sigma_{b}\sigma_{c}\sigma_{d}
\} \ .
\end{equation}
Now in the classical limit $ (\sigma^{4})^2 \rightarrow (\theta^{4})^2 = 0$ and
hence the axial-vector contribution vanishes. The tensor contribution will
survive because it is not surpressed by the commutation relations.
The general structure of Eq.(\ref{Srel}) is like that of the scalar density. We
have a sum of different particle and anti-particle contributions in which the
anti-particle contributions cancel. In the quantum-mechanical system we may
seperate between the magnetic dipole represented by ${\cal{S}}$ and the
spin-density represented by the axial-vector. In the classical limit however
this spindensity is killed by the commutation relations and only the magnetic
moment survives. If anti-particles are not present, i.e. in the
non-relativistic limit, the expectation values for spin and magnetic-dipole
densities are ofcourse proportional and so the breakdown of the Clifford
algebras anti-commutation relations causes no loss of physical information. Yet
in the classical limit nothing prevents us from going to extremely relativistic
energies where the appearance of anti-particles makes spin density and
dipole-moment density physically distinct. In this case the pseudo-classical
system still does not allow for such a d
spin density and magnetic-dipole density unless the anti-particles are
introduced by hand.
\par
\section{Conclusions}
\par
{}From the above elaborations we draw the following conclusions.
In the classical limit of a quantum transport theory for spin $\frac{1}{2}$
fermions spin can make its appearance in the form of a magnetic-dipole density.
In classical many-particle systems spin can become observable in a well-defined
manner. Basically the most general phase/spin space density will in the
classical
limit reduce to a sum of scalar, pseudo-scalar and vector contributions.
An axial-vector contribution, as is found in the spin-decomposition of the
non-relativistic Wigner function will not survive the classical limit.
This is due to the impossibillity of dynamically generating anti-particles in a
classical vacuum.
A tensor-contribution to the Wigner function, as is found in relativistic
quantum transport theory, need not vanish as $\hbar \rightarrow 0$.

\end{document}